\documentclass{jetpl}
\twocolumn
\lat
\usepackage{amssymb}
\usepackage{amsmath}
\usepackage{graphicx}
\usepackage{color}
\usepackage{bm}
\usepackage[colorlinks,linktocpage]{hyperref}
\hypersetup{colorlinks=true, citecolor=blue, linkcolor=blue, urlcolor=blue}
\usepackage[sort&compress,numbers]{natbib}
\usepackage{flushend}

\def\const{\mbox{const}}

\newcommand{\be}{\begin{equation}}
\newcommand{\ee}{\end{equation}}
\newcommand{\ba}{\begin{align}}
\newcommand{\ea}{\end{align}}
\newcommand{\bg}{\begin{gather}}
\newcommand{\eg}{\end{gather}}
\newcommand{\bseq}{\begin{subequations}}
\newcommand{\eseq}{\end{subequations}}

\renewcommand{\ln}{\mathop{\rm ln}\nolimits}

\makeatletter

\def\compoundrel#1\over#2{\mathpalette\compoundreL{{#1}\over{#2}}}
\def\compoundreL#1#2{\compoundREL#1#2}
\def\compoundREL#1#2\over#3{\mathrel
         {\vcenter{\hbox{$\m@th\buildrel{#1#2}\over{#1#3}$}}}}
\makeatother

\title{Numerical study of multiparticle production 
 in $\phi^4$ theory:\\comparison with analytic results}

\rtitle{Numerical study of multiparticle production in $\phi^4$ theory\ldots}
\sodtitle{Numerical study of multiparticle production in $\phi^4$ theory: 
comparison with analytic results}
\author{S.V. Demidov$\,^{a,b}$, B.R. Farkhtdinov$\,$\thanks{e-mail:
    farkhtdinov@phystech.edu}$^{a,b}$, D.G. Levkov$\,^{a,c}$}
\rauthor{S.\,V.\,Demidov, B.\,R.\,Farkhtdinov, and D.\,G.\, Levkov}
\sodauthor{Demidov, Farkhtdinov, Levkov}
\address{$^a$Institute for Nuclear Research of the Russian Academy of Sciences, 
Moscow 117312, Russia\\~\\
$^b$Moscow Institute of Physics and Technology, 
Dolgoprudny 141700, Russia\\~\\
$^c$Institute for Theoretical and Mathematical Physics,
Lomonosov Moscow State University, Moscow 119991, Russia}
\dates{3 November 2021}{*}

\abstract{We develop a numerical method to compute the probabilities
  of multiparticle production in weakly coupled scalar theories. Our
  technique is based on D.T. Son's semiclassical method of singular
  solutions. Applying it to the process $1 \to n$ in the unbroken
  four--dimensional $\lambda\phi^4$ theory, we reproduce the known
  results at ${1 \ll n \ll \lambda^{-1}}$.}

\begin{document}
\maketitle
%%%%%%%%%%%%%%%%%%%%%%%%%%%%%%%%%%%%%%
\section{Introduction}
\label{sec:intro}
It is well--known that perturbative expansion cannot be used to
calculate the amplitudes with large numbers of external legs $n\gtrsim
\lambda^{-1}$, where $\lambda$ is a small coupling
constant~\cite{Cornwall:1990hh, Goldberg:1990qk}. Indeed, resummation
of perturbative series in $\lambda\phi^4$ theory~\cite{Brown:1992ay,
  Voloshin:1992mz} indicates~\cite{Libanov:1994ug} that
multiparticle production occurs with exponentially small
probability at large $n$. Say, the
inclusive probability of creating $n \gg 1$ particles from one
off--shell particle equals
\be
\label{eq:0}
{\cal P}_{1\to n}\left(E\right) \equiv \sum_{f} |\langle
f; E,n|\hat{\cal S}\,\hat{\phi}(0) |0\rangle|^2 \propto {\rm
  e}^{F_{1\to n} / \lambda},
\ee
where $\hat{\phi}(0)$ creates an off--shell in-state,
$\hat{\cal S}$ is the S--matrix, the summation is performed over all final
states with energy $E$ and multiplicity $n$, and we ignored inessential
normalization factors and prefactors. Notably, the
suppression exponent $F_{1\to n} < 0$ in Eq.~(\ref{eq:0}) depends on
the combinations $\lambda n$ and $\lambda E$. 

A considerable revival of the interest in multiparticle processes
occurred recently~\cite{Voloshin:2017flq, Jaeckel:2018ipq,
Demidov:2018czx, Khoze:2018mey, Jaeckel:2018tdj, Schenk:2021yea}
when Ref.~\cite{Khoze:2017tjt} suggested that, contrary to
Eq.~(\ref{eq:0}), the cross section of multiple Higgs boson
production grows factorially at high energies. This ``Higgsplosion''
mechanism was subsequently criticised in~\cite{Demidov:2018czx,
  Belyaev:2018mtd, Monin:2018cbi, Dine:2020ybn}, so that now the
situation is far from being settled. It is clear that further development
of reliable methods for the calculation of multiparticle amplitudes is
required. 

Years ago, D.T. Son proposed~\cite{Son:1995wz} a general semiclassical
framework to calculate the multiparticle probabilities at $\lambda n
\sim O(1)$, see also~\cite{Khlebnikov:1992af, Diakonov:1993ha}. His 
technique is based on finding complex--valued singular solutions of
classical field equations with appropriate boundary
conditions. Despite being generic, this method was successfully 
applied only at $\lambda n \ll 1$ when semiclassical
configurations can be deduced from simplified semi--analytic
considerations.

In this Letter, we for the first time develop a complete numerical
implementation of the D.T.~Son's semiclassical method of singular
solutions. Our code computes the probability of the processes $1\to n$ in
four--dimensional unbroken $\lambda \phi^4$ theory at arbitrary
$\lambda n \sim 1$ and $\lambda E \sim O(m)$, where $m$ is the
particle mass.  As an initial step, we present here numerical results
at $\lambda n \ll 1$ and demonstrate that they agree with  
predictions of the perturbation theory.

%%%%%%%%%%%%%%%%%%%%%%%%%%%%%%%%%%%%%%%%%
\section{Semiclassical method for multiparticle
  production}
\label{sec:semicl-meth-num-impl}
In this Section, we review the method of~\cite{Son:1995wz} in application to a 
weakly coupled $(3+1)$--dimensional scalar field theory with the action, 
\begin{equation}
\label{eq:1.1}
S = \frac{1}{2\lambda}\int d^4 x \left(\,  -  \phi \Box \phi - \phi^2 
- \phi^4/2 \, \right)\;.
\end{equation}
Here $\lambda \ll 1$ is the coupling constant that simultaneously plays
the role of a semiclassical parameter and we work in units of the
field mass $m=1$. 

It is convenient to introduce the current $J$   
\begin{equation}
\label{eq:1.2}
{\mathcal{P}}_J(E,n)=\sum_{f} |\langle
f;n,E|\hat{\cal S}\,{\rm e}^{-
  J\hat{\phi}(0)/\lambda }|0\rangle|^2,  
\end{equation}
so that the probability~\eqref{eq:0} equals 
\begin{equation}
  \label{eq:1}
{\mathcal{P}_{1 \to n} = \lambda^2 \, \lim\limits_{J \to 0}
  {\mathcal{P}}_J/J^2}\;.
\end{equation}
In~\cite{Son:1995wz}, the quantity~\eqref{eq:1.2} was represented
as a path integral which is saturated at small $\lambda$ by a
complex-valued saddle--point configuration $\phi(t,\, {\bm
  x})$. The saddle--point conditions for the latter include a
classical field equation with the source term,
\begin{equation}
\label{eq:1.3}
\Box\, \phi(x) + \phi(x) + \phi^3(x) = i J\delta^{(4)}(x),
\end{equation} 
and certain boundary conditions. In particular, the semiclassical
configuration should contain only the  positive--frequency
part in the infinite past,
\be
\label{eq:1.4}
\phi \to \int d^3 {\bm k} \; {\rm e}^{-i{\bm k}{\bm x}+i\omega_{\bm
    k}t}\;  a_{\bm k}\qquad \mbox{as} \qquad t\to -\infty\;,
\ee
where  $a_{\bm{k}}$ are arbitrary and
  $\omega_{\bm{k}}= ({\bm k}^2+1)^{1/2}$. At large positive times $t\to
+ \infty$ the solution is expected to linearize:
\begin{equation}
\label{eq:1.5}
  \phi \to \int \frac{d^3{\bm k}\;  {\rm
      e}^{i{\bm{kx}}}}{(2\pi)^{3/2}\sqrt{2\omega_{\bm k}}}
  \left(\, f_{\bm k}{\rm e}^{-i\omega_{\bm k}t}  
 + g_{-{\bm k}}^*{\rm e}^{i\omega_{\bm k}t}\, \right).
\end{equation}
Saddle--point equations in this case relate 
the positive-- and negative--frequency components of $\phi$, 
\be
\label{eq:1.6}
f_{\bm{k}} = {\rm e}^{-\theta + 2\omega_{\bm{k}}T}\, g_{\bm{k}}\,,
\ee
where $T$ and $\theta$ have the sense of Lagrange multipliers
appearing due to the fixation of energy $E$ and final--state
multiplicity $n$. The latter quantities are given by the standard 
expressions,
\begin{equation}
\label{eq:1.7}
\lambda E= \int d^3 {\bm k}\, \omega_{\bm{k}}\,  f_{\bm{k}}
g_{\bm{k}}^* \;, \quad \lambda n = \int d^3 {\bm{k}}\, f_{\bm{k}} g_{\bm{k}}^*\, .
\end{equation}
In what follows we parametrise the solutions with the rescaled
multiplicity $\lambda n$ and kinetic energy per particle~${\varepsilon
\equiv E/n-1}$.

Once the semiclassical equations are solved, one finds the
probability~\eqref{eq:0} by taking the limit
\begin{equation}
\label{eq:1.9}
{\cal P}_{1\to n} \approx \lim_{J \to 0} {\cal P}_J  \approx \lim_{J
  \to 0} \mathrm{e}^{F_J/\lambda}\,, 
\end{equation}
where the prefactors are ignored and  $F_J$ is the value of the
functional
\begin{equation}
\label{eq:1.10}
F_J/\lambda = 2ET-
n\theta-2{\rm   Im}\,{S} - 2 J\, {\rm Re}\,\phi(0)/\lambda\;,
\end{equation}
computed on the saddle-point solution $\phi(t,\,
\bm{x})$. 

It is important that the method of~\cite{Son:1995wz} involves a
nontrivial assumption that the suppression exponent in
Eq.~\eqref{eq:0} is {\it universal}, i.e.\ does not depend on the  
details of a few--particle initial state~\cite{Rubakov:1992ec,
  Tinyakov:1991fn, Bonini:1999kj, Levkov:2008csa}. In particular, the
exponent is not sensitive to the choice of the source term in
Eq.~(\ref{eq:1.2}). However, in any case, the semiclassical solutions
become singular at $t=0$ in the limit $J\to 0$, since their energies are equal
to zero and $E$ at $t<0$ and $t>0$,  respectively~--- see
Eqs.~\eqref{eq:1.4}, \eqref{eq:1.7},  and~\cite{Libanov:1997nt}.   

In the previous studies~\cite{Son:1995wz, Bezrukov:1995qh, Bezrukov:1998mei} the
semiclassical solutions were found analytically at small $\lambda n$
and $\varepsilon$. It was shown that the semiclassical exponent
$F_{1\to n}$ agrees with the one--loop perturbative results of
Refs.~\cite{Voloshin:1992mz, Voloshin:1992nu} in that region.

%%%%%%%%%%%%%%%%%%%%%%%%%%%%%%%%%%%%%%%%%%%%%
\section{Numerical results}
\label{sec:num-res}
\begin{figure}[!htb]
  \centerline{\includegraphics{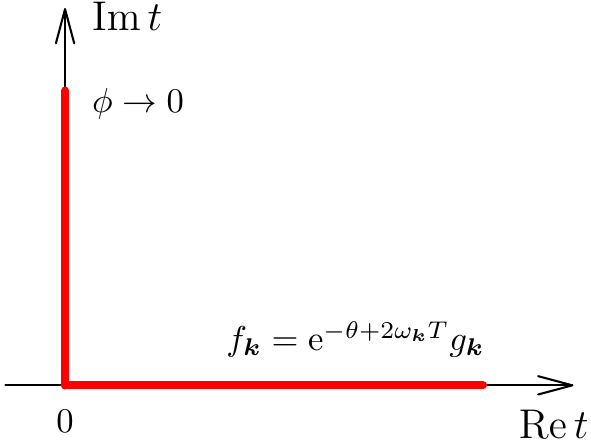}}
  \caption{Figure 1. Contour in complex time for the semiclassical
    boundary value problem~\eqref{eq:1.3}--\eqref{eq:1.7}.\label{fig_cont}}
\end{figure}

Let us outline the numerical method for solving the boundary
problem~\eqref{eq:1.3}--\eqref{eq:1.7} at arbitrary $\lambda   
n$ and $\varepsilon$. We analytically continue the solution to the
complex time contour in Fig.\ref{fig_cont}.
Then Feynman initial condition~\eqref{eq:1.4}
takes the form,
\be
\label{eq:2.0}
\phi(t,\, {\bm x})\to 0\qquad {\rm as}\qquad {\rm Im}\,t\to +\infty\,.
\ee
Besides, we regularize the source replacing 
\begin{equation}
\label{eq:2.1}
J\delta^{(4)}(x)\to j\,  {\rm e}^{-{\bm x}^2/2\sigma^2}\delta(t)\, ,
\end{equation} 
where $j$ and $\sigma$ will be sent to zero simultaneously. Next, we
substitute the spherically--symmetric Ansatz $\phi = \phi(t,\, r)$
into the equation~\eqref{eq:1.3} and boundary
conditions~\eqref{eq:1.4}--\eqref{eq:1.6} and discretize the resulting
system on the rectangular space--time lattice with sites $(t_j,\,
r_k)$. The discrete problem is then solved using the
Newton--Raphson method~\cite{NR}. 

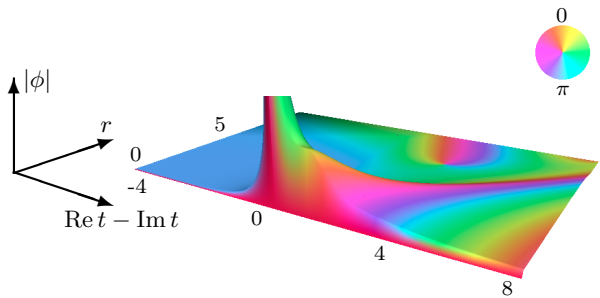
\begin{figure}[!thb]
\unitlength 0.6mm
\centerline{\begin{picture}(20,90)(0,-20)
    \thicklines
\put(-9.0,4.0){\includegraphics[width=105\unitlength,angle=0]{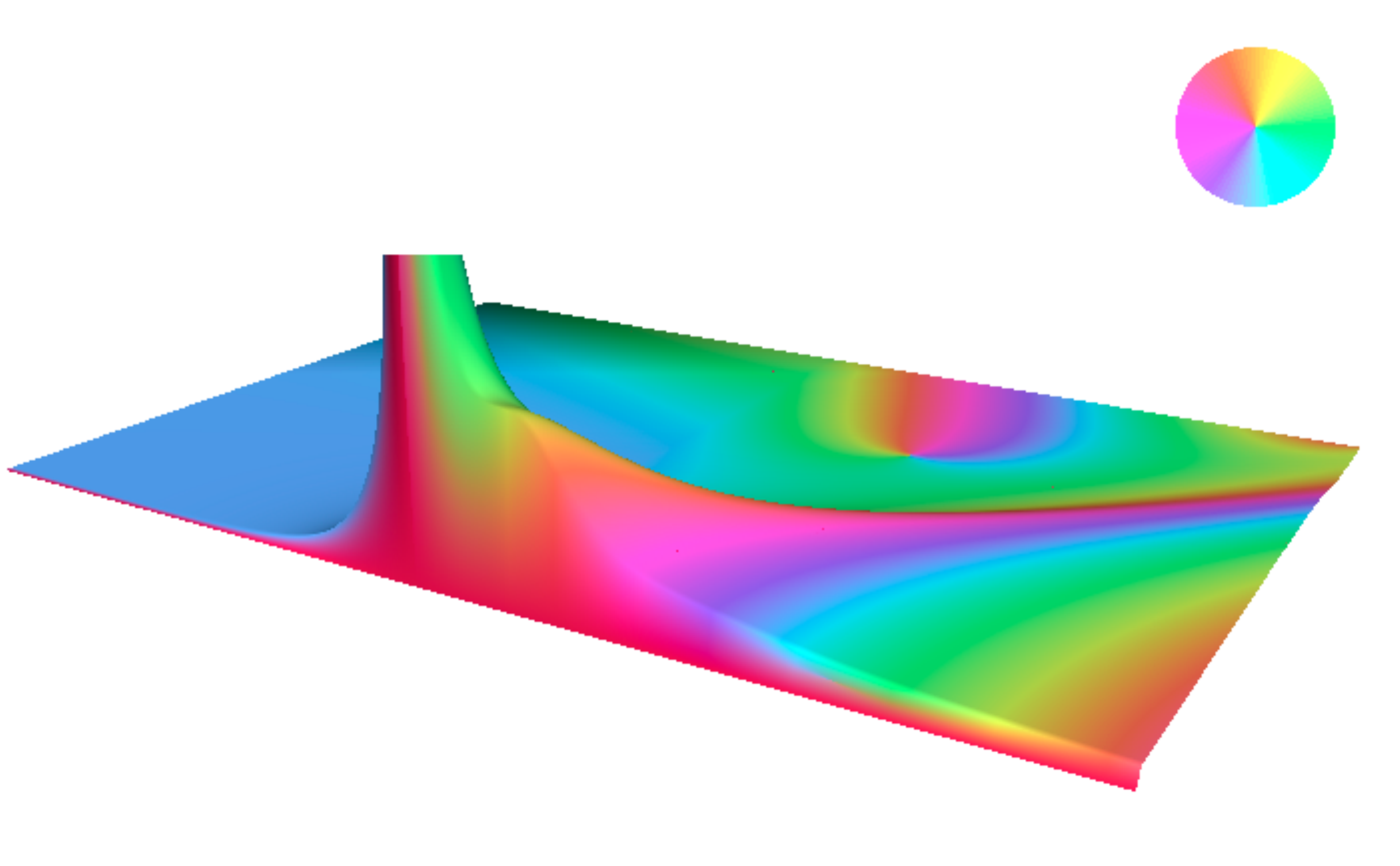}}
\put(-35,33){\line(3,-1){20.0}}
\put(-14.0,25.9){\vector(3,-1){1}}
\put(-35,33){\line(3,1){20.0}}
\put(-14,40){\vector(3,1){1}}
\put(-35,33){\line(0,1){20.0}}
\put(-35,53){\vector(0,1){1}}
\put(-24,21){\small ${\rm Re}\,t-{\rm Im}\,t$}
\put(-16,42){\small $r$}
\put(-33,52){\small $|\phi|$}
\put(-10.0,28.4){\footnotesize{-4}}
\put(17.0,21.4){\footnotesize{0}}
\put(44.0,13.7){\footnotesize{4}}
\put(72.0,6.0){\footnotesize{8}}
\put(-9.5,35.5){\footnotesize{0}}
\put(9.0,42.1){\footnotesize{5}}
\put(83.9,66.2){\footnotesize{0}}
\put(83.9,50.1){\footnotesize{$\pi$}}
\end{picture}\hspace{2cm}}
\vspace{-15mm}
\caption{Figure 2. Semiclassical solution $\phi(t,\, r)$ with
  parameters ${\varepsilon \approx 1.35}$, ${\lambda n\approx 0.38}$,
  $j=0.3$, and ${\sigma\approx 0.1}$. Colour shows complex phase
  of $\phi$. 
\label{fig_solution}} 
\end{figure}

Changing $T$, $\theta$, $j$, and $\sigma$ in small steps, we find all
regularised numerical solutions. The respective exponents
$F_J(\varepsilon,\lambda n)$ are computed by performing integration in 
Eq.~\eqref{eq:1.10}. An example of our semiclassical configuration
$\phi (t,\, r)$ is given in Fig.~\ref{fig_solution}.
One observes a high and narrow peak at $t= r=0$ where the source is
located. The solution becomes singular at this point in the limit
$j,\sigma \to 0$. The outgoing waves represent the final--state 
particles emanating from the source.

The semiclassical expression~(\ref{eq:1.9}) for the probability
${\cal P}_{1\to n}\approx \exp\{F_{1\to n}/\lambda\}$ includes
the limit $j,\, \sigma\to 0$. Notably, the saddle--point
configurations cannot be directly computed at $j = 0$, as they
are singular. We, therefore, perform a polynomial extrapolation of $F_J$
to $j= 0 $ keeping $j/\sigma = \const$, cf.\ Eq.~(\ref{eq:1}). The
technical details of this procedure  will be presented
elsewhere~\cite{in_prep}.

To verify the numerical technique, we compare the multiparticle
probability with the known perturbative results at small $\lambda
n$. In this case, the semiclassical exponent  $F_{1 \to n}$ has 
the form~\cite{Son:1995wz}, 
\begin{equation}
\label{eq:2.2}
F_{1\to n}= \lambda n \ln \left( \frac{\lambda
  n}{16} \right) - \lambda n + \lambda n f(\varepsilon) + O(\lambda^2 n^2) \,,
\end{equation}
where $\varepsilon \simeq O(1)$ and the function $f(\varepsilon)$ is
unknown. The first three terms in Eq.~\eqref{eq:2.2} and, notably,
$f(\varepsilon)$, can be extracted from tree--level diagrams. The
latter calculation was performed numerically
in~\cite{Bezrukov:1998mei} for arbitrary $\varepsilon$. It is worth
noting that their result at $\varepsilon \le 10$ almost saturates the
simpler $O(4)$ bound of Ref.~\cite{Bezrukov:1995ta}. Below we use the
latter for comparison.

\begin{figure}[!htb]
  \unitlength=1mm
  \centerline{\begin{picture}(86,65)
      \put(0,0){\includegraphics{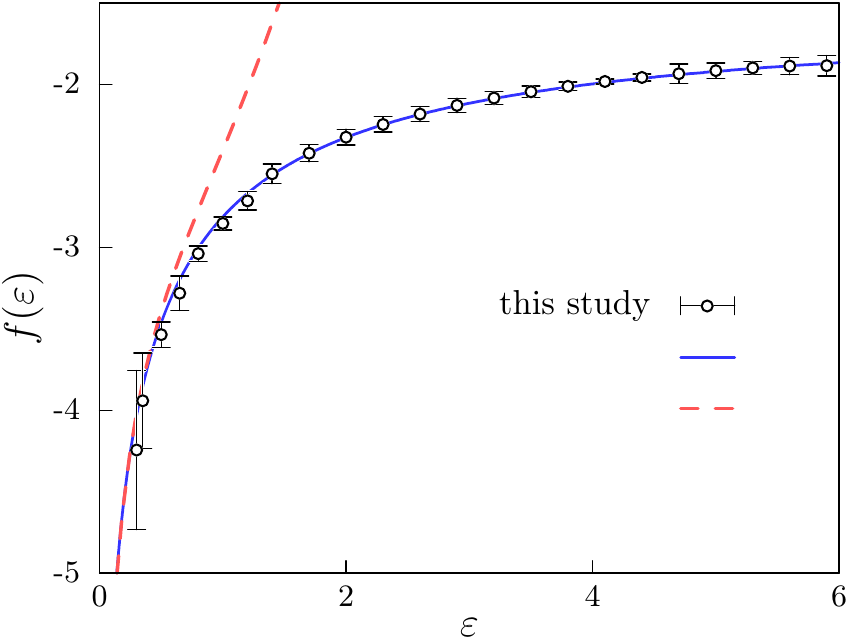}}
      \put(26.2,22.7){low--$\varepsilon$ expansion,
        Eq.~\eqref{eq:2.3}}
      \put(29.8,27.7){$O(4)$ estimate, Ref.~\cite{Bezrukov:1995ta}}
    \end{picture}}
    \caption{Figure 3. Numerical values of $f(\varepsilon)$ (circles
      with errorbars) compared with the tree--level results:
      $O(4)$--symmetric estimate of Refs.~\cite{Bezrukov:1995ta,
        Bezrukov:1998mei} (solid line) and low--$\varepsilon$
      expansion~\eqref{eq:2.3} (dashed). \label{fig_result}}
\end{figure}

Our results for $f(\varepsilon)$ are shown in Fig.~\ref{fig_result} 
by circles with errorbars which represent inaccuracies of
the extrapolation $j \to 0$. These numerical data cover a 
finite energy range ${0.3 \le \varepsilon \le 6}$. At smaller
$\varepsilon$, the saddle--point solutions become nonrelativistic and fail to fit
into the spacetime volume available for computations. At $\varepsilon
\ge 6$ the solutions include higher--frequency waves which cannot be
resolved.

Notably, the  data points in Fig.~\ref{fig_result} are consistent
with the tree--level results previously obtained in the literature.
They coincide with the $O(4)$ curve of Ref.~\cite{Bezrukov:1995ta}
(solid line), as they should. Besides, both numerical graphs approach
the low--$\varepsilon$ asymptotics of $f(\varepsilon)$, 
\begin{multline}
f = \frac{3}{2} \ln{\frac{\varepsilon}{3\pi}}
+\frac32 - \frac{17}{12}\, \varepsilon \\
+ \frac{1327-96\pi^2}{432}\, \varepsilon^2 +  O(\varepsilon^3)\,,
\label{eq:2.3}
\end{multline} 
that was evaluated analytically in~\cite{Bezrukov:1995qh}.

%%%%%%%%%%%%%%%%%%%%%%%%%%%%%%%%%%%%%%%%%%
\section{Conclusions}
\label{sec:concl}
We developed a numerical method to compute semiclassically the
probabilities of $n$--particle production in scalar theories in the
regime $n \to \infty$ and $\lambda n=$~fixed, where $\lambda$ is a
small coupling constant. We illustrated the method by performing
explicit calculations in a four--dimensional unbroken $\lambda \phi^4$
model. At $\lambda n \ll 1$ our numerical data for the probability
agree with the tree--level results obtained previously in the
literature. But notably, our technique is also applicable at~${\lambda n
  \sim O(1)}$.

This work is supported by the RFBR, grant \textnumero~20-32-90013. Numerical
calculations were performed on the Computational cluster of the
Theoretical Division of INR RAS.  

%%%%%%%%%%%%%%%%%%%%%%%%%%%%%%%%%%%%%
\bibliographystyle{apsrev4-1}
\bibliography{note_numer}

\end{document}